\documentclass[referee]{aa}

\usepackage{graphics}
\usepackage{psfig}
\begin{document}

\title{ A Method for Simultaneous Determination of Metallicity
and Reddening of a Globular Cluster based on the $V vs. (B-R)$
Color-Magnitude Diagram}
\author{A.N.\,Gerashchenko,}
 \offprints{A.N.Gerashchenko, The Central Astronomical Observatory,
Pulkovo, St.\,Petersburg, Russia. email: ger@gao.spb.ru}

\institute{The Central Astronomical Observatory of RAS at Pulkovo,
St.\,Petersburg, Russia}

\date{ }

\authorrunning{Gerashchenko}

\titlerunning{ A Method for Determination of Metallicity
and Reddening}

\abstract{ We describe a method by which the abundance of metals
and the reddening of globular clusters are simultaneously derived.
The method is based on the analysis of the shape of the red giant
branch and V magnitude of the horizontal branch on the $V vs.
(B-R)$ color-magnitude diagram for a cluster. The application of
our technique to the photometry of the globular cluster NGC\,7006
yields the metallicity $[Fe/H] = -\,1.78 \pm0.11$ and the
reddening $E_{B-R} = 0.^{m}25 \pm0.^{m}02$.

}

\maketitle

\section{Introduction}

Metallicity and interstellar absorption are the most important
characteristics in studies of globular clusters: the first
specifies their basic properties, while the second, on the
contrary, makes it difficult to determine them. A number of
studies have been devoted to the relationship between parameters
of the giants branch (GB) on the color - magnitude diagram and
metallicity (for example, Carretta and Bragaglia (1998),
Sarajedini and Layden (1997), Zinn and West (1984), Sandage
(1982)). However, to determine most of GB parameters, the
interstellar absorption should be known. Therefore, it is of great
importance to develop a technique for simultaneous determination
of both the reddening and the GB parameters. Such techniques were
suggested by Sarajedini and Norris (1994) and Ferraro et al.
(1999) for $V vs.(B-V)$, by Sarajedini (1994) for $V vs.(V-I)$
diagrams. Here, we describe a technique by which the metallicity
and the interstellar absorption can be simultaneously derived and
which is based on the analysis of $V vs.(B-R)$ diagrams for
globular clusters.

\section{The observational data}

To develop our technique, a homogeneous set of observations of
globular clusters in $B, V, R $ passbands, distributed
homogeneously over broad intervals of metallicities and reddenings
is needed. We used the $B, V, R_c, I_c $ Johnson-Cousins
photometry of stars in globular cluster carried out by Alcaino et
al., which satisfies the above requirements. Table 1 contains the
basic parameters of the observed clusters: the name of each
cluster, the metallicity in Zinn's scale (see Armandroff and Zinn
(1988), Zinn (1985); Zinn and West (1984); this scale, despite
some drawbacks, represents the most complete and homogeneous
database available), the reddening $E_{B-V}$ and the visual
magnitude of horizontal branches (HB) $V_{HB}$ taken from the
Harris catalogue (1996, version 2003); the references to published
$BVR$ and $BV$ photometrical observations. The $BVR$ magnitudes of
stars in three globular clusters (NGC\,5286, NGC\,6541 and
NGC\,6723) were kindly presented by N.N.Samus'.

\begin{table}[!hb]
\begin{center}
{ Table 1. The basic parameters of the globular clusters\\}
\vspace{0.2cm}
\begin{tabular}{|c|c|c|l|c|c|}
\hline
  NGC & $[Fe/H]$  & $E_{B-V}$& $V_{HB}$ &$BVR$ references &$BV$ references\\
\hline
   1  &    2   &   3  &  4  &  5  &  6  \\
\hline

 1261 &  -1.29  & 0.01  & 16.7  &  1 &13\\

1904 &  -1.68  & 0.01  & 16.15 &  2 & 14 \\

2298 & -1.81   & 0.14  & 16.11 &  3 &    \\

2808 & -1.71   & 0.22  & 16.22 &  4 & 15 \\

3201 & -1.56   & 0.23  & 14.77 &  5 & 16 \\

4590 & -2.09   & 0.05  & 15.68 &  6 & 17 \\

5286 &  -1.79  & 0.24  & 16.50 &  7 &    \\

6121 &  -1.28  & 0.36  & 13.45 &  8 & 18 \\

6362 & -1.08   & 0.09  & 15.33 &  9 & 19 \\

6541 & -1.83   & 0.14  & 15.20 & 10 &    \\

6723 & -1.09   & 0.05  & 15.48 & 11 &    \\

6809 & -1.82   & 0.08  & 14.4  & 12 & 20 \\
 \hline

\end{tabular}
\end{center}
\end{table}

Footnote to Table 1:{\small
 the numbers in columns 5 and 6 denote
references to published $BVR$ and $BV$ photometrical observations:
 1 - Alcaino et al. (1992a), 2 - Alcaino et al. (1994), 3 - Alcaino and
Liller (1986a), Alcaino et al. (1990a), 4 - Alcaino et al.
(1990b), 5 - Alcaino et al. (1989), 6 - Alcaino et al. (1990c), 7
- Samus' et al. (1995), 8 - Alcaino and Liller (1984), 9 - Alcaino
and Liller (1986b), 10 - Alcaino et al. (1997), 11 - Alcaino et
al. (1999), 12 - Alcaino et al. (1992b), 13 - Ferraro et al.
(1993), Alcaino (1979), 14 - Ferraro et al. (1992), 15 - Ferraro
et al. (1990), 16 - Lee (1977a), 17 - Walker (1994), 18 - Lee
(1977b), 19 - Alcaino et al. (1972), 20 - Lee (1977c)}

\section{The $V vs.(B-V)$ color-magnitude diagrams.}

In order to estimate the quality of $B,V,R_c,I_c$ observations
made by Alcaino et al. and their adequacy to our purposes,
initially we constructed $V vs.(B-V)$ diagrams for the clusters.
For most of the clusters, the bright part of the diagrams appeared
to be poorly populated, which is explained by the purpose of the
observations: determination of cluster's ages and, hence,
obtaining stellar magnitudes for sufficiently faint stars. In
turn, it may result in unreliable representation of the GB by some
function, which in our case is a cubic polynomial. Therefore, for
each cluster from Table 1 we selected the most accurate
photometric $BV$ data available in the literature (see references
to them in column 6 of Table 1), which represent the total GB.
Comparing this set of observations with that obtained by Alcaino
et al. and introducing corrections where necessary, we reduced it
to the to system of $B,V,R,I$ magnitudes of the latter set.

In the $V vs. (V-B)$ color - magnitude diagrams constructed
thereby, the GB for each cluster was represented by a 3-rd order
curve, from which the following GB parameters were determined: the
GB slope - $S_{2.0}$ (Sandage and Wallerstein, 1960) and $S_{2.5}$
(Sarajedini and Layden, 1997); the height of the GB above the HB
level - $\Delta{V_{1.1}}$, $\Delta{V_{1.2}}$ (Sarajedini and
Layden, 1997), and $\Delta{V_{1.4}}$ (Sandage and Wallerstein,
1960) in the points with $(B-V)_{o} = 1.^{m}1$, $1.^{m}2$, and
$1.^{m}4$, respectively; the intrinsic $(B-V)_{o}$ color of the GB
$(B-V)_{o,g}$, measured at the HB level. In the calculations for
$(B-V)_{o,g}$, $\Delta{V_{1.1}}$, $\Delta{V_{1.2}}$, and
$\Delta{V_{1.4}}$, the reddening was taken in accordance with
Table 1.

The calculated GB parameters were calibrated with respect to
metallicity $[Fe/H]$ with the use of Table 1 data. The correlation
(r) of $S_{2.0}$ or $S_{2.5}$ with metallicity is high (see
Fig.1):

 $$[Fe/H]=0.64-0.34S_{2.0};  r = 0.92;  \sigma = 0.14; n =
 11,  \eqno (1)$$

 $$[Fe/H]=0.46-0.37S_{2.5};  r = 0.92;  \sigma = 0.14; n = 11,
  \eqno (2)$$

where $\sigma$ is the rms error, n -- the number of the clusters.
The cluster NGC\,6541 which have the largest deviation in fig 1
was excluded from the calculations for the expressions (1), (2)
and for the subsequent "metallicity - $\Delta{V}$" relations.

 \begin{figure} [!h]
\centering{
 \vbox{\psfig{figure=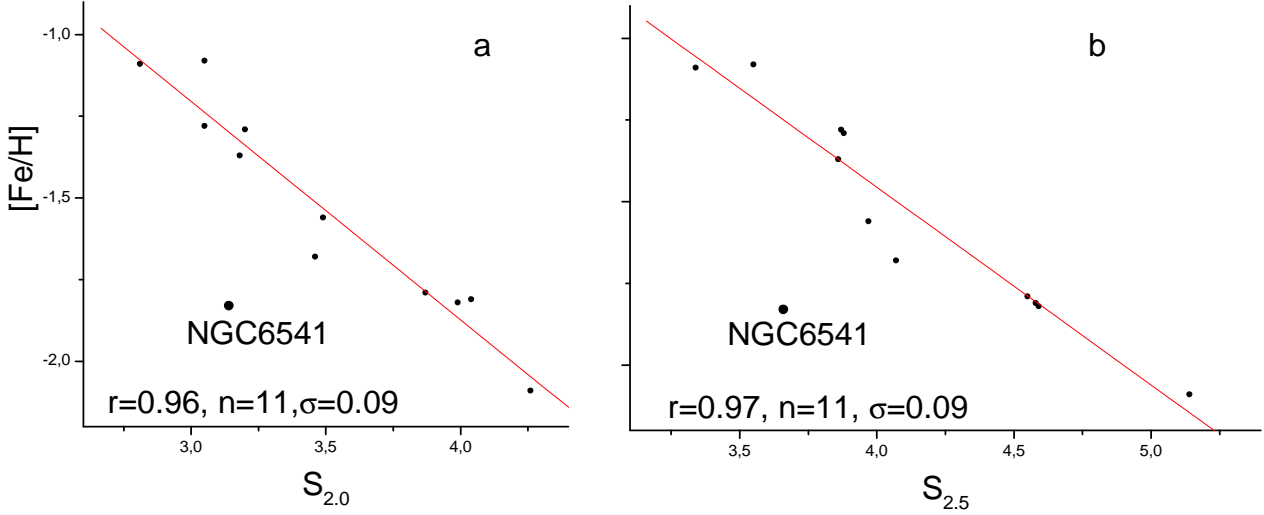,width=17cm}}\par }
 \caption[]{Calibration of the GB parameters derived from the $V vs. (B-V)$
 globular cluster diagrams with respect $[Fe/H]$: (a) - $S_{2.0}$,
 (b) - $S_{2.5}$. The solid lines are the least-squares fits to the data.
 The number of the clusters (n) used to compute each relation is reported
 together with the standard deviations ($\sigma$) of the data.}
 \end{figure}

The correlation between the metallicity and the parameters
$(B-V)_{o, g}$, $\Delta{V_{1.1}}$, $\Delta{V_{1.2}}$, and
$\Delta{V_{1.4}}$A, which depend on interstellar absorption, is
slightly worse, within $(86-89)\%$. The possible reason for that
is some inaccuracy in the adopted values of reddening for the
globular clusters that display the largest scattering on
calibration curves. Therefore, the values $E_{B-V}$ for these
clusters were redefined. Staying true to Sarajedini (1994), we
determine the true reddening of a cluster as that which value
yields the numerical decision for the system of equations
analogous to (3) - (6).

$$[Fe/H] = a1+b1(B-V)_{o,g}                       \eqno(3)$$

$$[Fe/H] = a2+b2\Delta{V_{1.1}}+ c2\Delta{V_{1.1}^2} \eqno (4)$$

$$[Fe/H] = a3+b3\Delta{V_{1.2}}+ c3\Delta{V_{1.2}^2} \eqno (5)$$

$$[Fe/H] = a4+b4\Delta{V_{1.4}}+ c4\Delta{V_{1.4}^2} \eqno (6) $$

As a result, for 5 clusters the reddening $E_{B-V}$ is different
from that given in Table 1: NGC\,2298 - $0.^{m}21$, NGC\,3201 -
$0.^{m}18$, NGC\,5286 - $0.^{m}28$, NGC\,6121 - $0.^{m}40$,
NGC\,6809 - $0.^{m}16$. These values, basically, are within the
limits in which $E_{B-V}$ are determined by different authors. For
example, $E_{B-V}$ received by us exactly coincide with the values
obtained by Alcaino and Liller (1986a) for NGC\,2298 and Lee
(1977b) for NGC\,6121. However, for two clusters the system of
equations (3) - (6) has the solution only for following values
$E_(B-V)$: $E_(B-V) \leq 0.^{m}18$ for NGC\,3201 and $E_(B-V) \geq
0.^{m}16$ for NGC\,6809. After the acceptance of the new values
for color excesses, the system of equations (3) - (6) for clusters
from Table 1 (with the exception of NGC\,6541 for equations (8) -
(10)) has the form (fig 2):

$$ [Fe/H] = -5.60+5.18(B-V)_{o,g}; r=0.95; \sigma=0.10; n=12,
\eqno
 (7)$$

 $$[Fe/H] = -1.00+0.32\Delta{V_{1.1}}-0.28\Delta{V_{1.1}^2};
  r=0.94; \sigma=0.09; n=11, \eqno (8)$$

 $$[Fe/H] = -0.98+0.30\Delta{V_{1.2}}-0.22\Delta{V_{1.2}^2};
  r=0.94; \sigma=0.09; n=11, \eqno (9)$$

 $$[Fe/H] = 1.68-1.45\Delta{V_{1.4}}-0.11\Delta{V_{1.4}^2};
  r=0.88; \sigma=0.10; n=11 \eqno (10)$$

 \begin{figure} [!h]
 \centering{
 \vbox{\psfig{figure=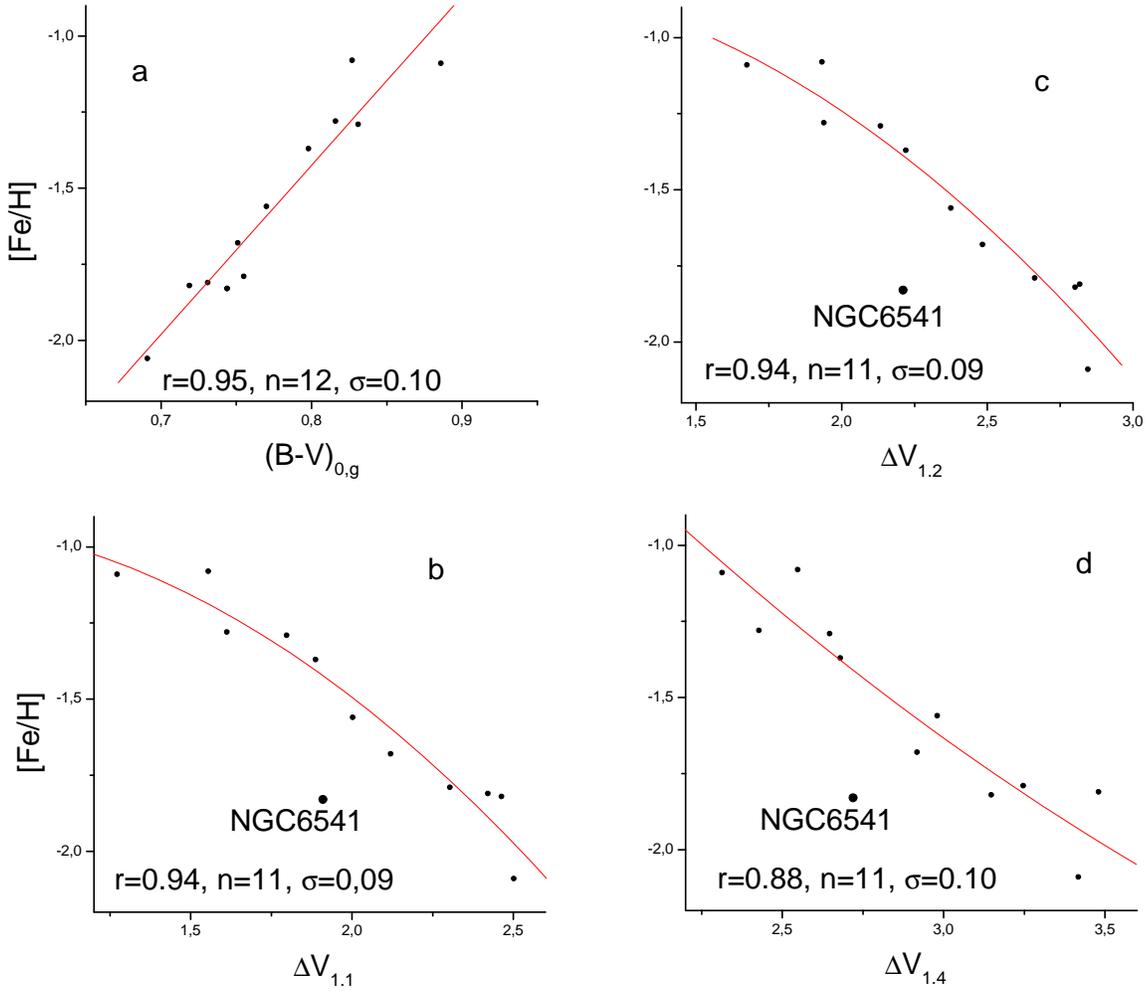,width=16cm}}\par }
 \caption[]{Calibration of the GB parameters derived from the $V vs. (B-V)$
 globular cluster diagrams with respect $[Fe/H]$: a - $(B-V)_{o,g}$,
 b - $\Delta{V_{1.1}}$, c - $\Delta{V_{1.2}}$, d - $\Delta{V_{1.4}}$.
 Text is the same as in fig.1.}
 \end{figure}

The correlation for $\Delta{V_{1.4}}$ deteriorates due to the fact
that in metal-poor clusters the GB does not reach colors
$(B-V)_{o}\geq1.4$ and the extrapolation for these colors is
unreliable. For metal-rich clusters, this part of the
color-magnitude diagram is populated by red variable stars, which
also makes the interpolation unreliable.

When the obtained equations are compared with those published in
the literature, it is apparent that they agree qualitatively. A
direct comparison is possible only with respect to the equation
(7). The coefficients of this equation correspond to those
obtained by Sarajedini and Norris (1994) within the errors of
determination. For the other equations, no direct comparison is
impossible, either because of different scales of metallicity or
since the in other studies equations of other degrees are
presented.

From the above we can conclude that the observational data obtained
by Alcaino with co-authors is consistent with the goals of our study,
after the modifications described.

\section{The $V vs. (B-R)$ color - magnitude diagrams.}

As it was noted above, the data of Alcaino's observations poorly
represent the brightest part of the GB on the $V vs. (B-V)$ color
- magnitude diagram. It also refers to the $V vs. (B-R)$ or $V
vs.(V-R)$ diagrams. Therefore, the construction of these diagrams
and determination for the GB parameters on them should be carried
out as follows. From the data obtained by Alcaino et al., we
derive relations between $(B-V)$ and $(B-R)$, and also between
$(B- V)$ and $(V-R)$ colors for each cluster. The correlation of
$(B-V)$ and $(B-R)$ colors is high $(>99\%)$, substantially
exceeding that in the case of $(B-V)$ and $(V-R)$ colors, which
justifies the selection of the $V vs. (B- R)$ diagram for our
study. With the use of these relations, we transform the $V
vs.(B-V)$ diagrams described in the previous paragraph into $V
vs.(B-R)$. When these diagrams are compared with the data obtained
by Alcaino et al., no systematic deviations between them are seen.
The GB parameters are determined from $V vs. (B-R)$ diagrams in a
way similar to that used for $V vs. (B-V)$ diagrams.

The relations between these parameters and metallicity have been
derived using Table 1 and the $E_{B-V}$ values for some clusters,
specified in the previous paragraph. Color excesses $E_{B-V}$ are
transformed to $E_{B-R}$ through the Grebel and Roberts (1995)
ratio: $E_{B-R}/E_{B-V} $ = 1.62.

 \begin{figure} [!h]
 \centering{
\vbox{\psfig{figure=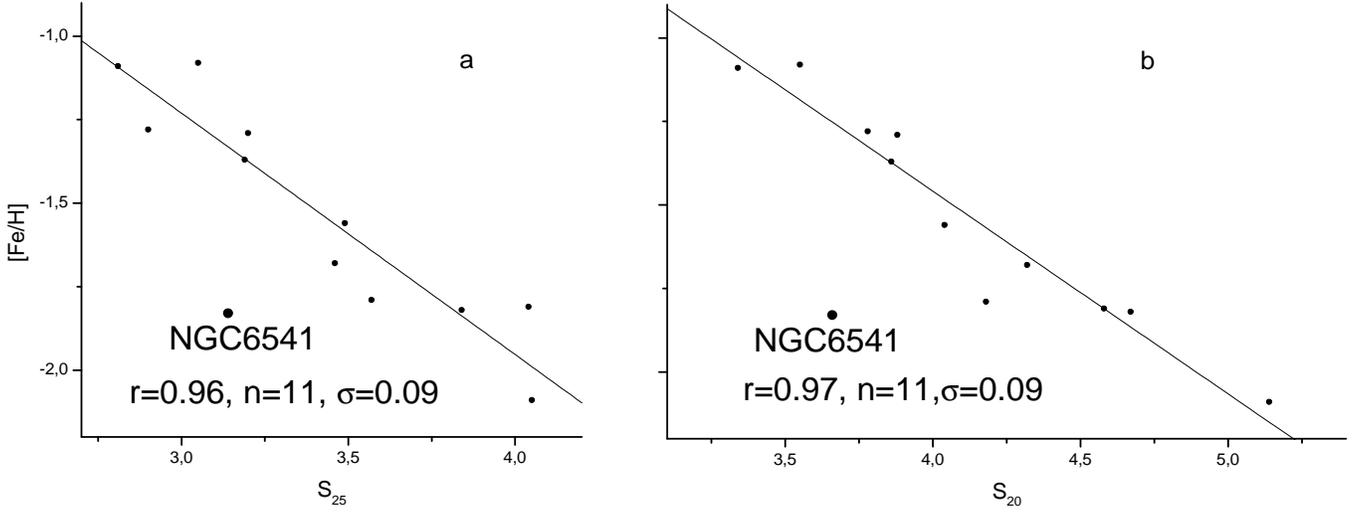,width=18cm}}\par }
  \caption[]{The same as in fig.1, but for diagrams  $V vs.(B-R)$}
    \end{figure}

The equations that connect the GB parameters and metallicity of
the clusters (fig 3 and fig 4) can be used to determine the
metallicity $[Fe/H]$ and color excesses $E_{B-R}$:

$$
 [Fe/H] = 0.97-0.61S_{2.0}; r = 0.96;  \sigma = 0.09; n = 11, \eqno
 (11)$$

$$[Fe/H] = 0.88-0.71S_{2.5}; r = 0.97; \sigma = 0.09; n = 11,
\eqno (12)$$

$$(B-R)_{o,g} = 1.64 + 0.27[Fe/H]; r = 0.96; \sigma=0.03;  n=12,
\eqno (13)$$

$$[Fe/H] = -0.82+0.01\Delta{V_{1.6}}-0.23\Delta{V_{1.6}^2};
r=0.92; \sigma=0.10; n=11,  \eqno (14)$$

$$[Fe/H] = -0.59+0.01\Delta{V_{1.7}}-0.18\Delta{V_{1.7}^2};
r=0.92; \sigma=0.11; n=11, \eqno (15)$$

$$[Fe/H] = -0.63+0.15\Delta{V_{1.9}}-0.21\Delta{V_{1.9}^2};
r=0.93; \sigma=0.10; n=11, \eqno (16)$$

$$[Fe/H] = -5.96+3.64(B-R)_{o,g}; r= 0.95; \sigma = 0.11; n = 12
 \eqno (17) $$

 \begin{figure} [!h]
 \centering{
\vbox{\psfig{figure=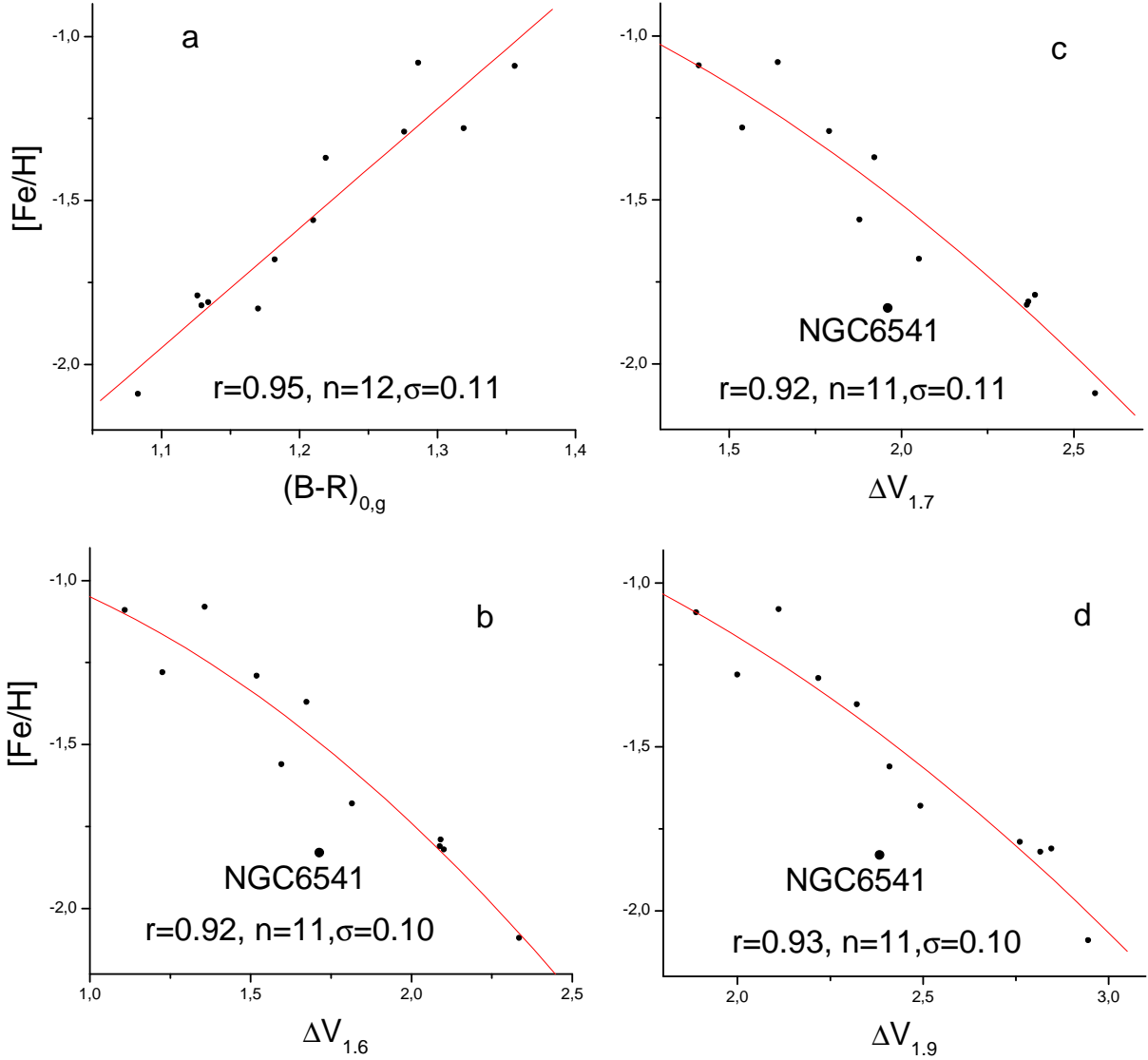,width=17cm}}\par }
   \caption[]{The same as in fig.2, but for diagrams  $V vs.(B-R)$}
      \end{figure}

When the equations (1,2,7-10) are compared with (11-17), their
correlation coefficients and rms show that the accuracy of
determination of metallicity and color excesses through $V vs.
(B-R)$ diagrams is not worse than through $V vs.(B-V)$ diagrams.

Below, we consider the application of the equations (11-17) to
simultaneous determination for the metallicity [Fe/H] and color
excesses $E_{B-R}$ for the globular cluster NGC\,7006.

\section{The application of the method to NGC\,7006}

The observational data for the cluster NGC\,7006, used to apply
our method, were obtained with the Zeis-1000 Telescope at SAO RAS
equipped with the K-585 CCD-detector in $B, V, R_c$ Johnson -
Cousins passbands (Gerashchenko, 2007).

NGC\,7006 is a cluster of intermediate metallicity; its values,
determined by various methods, vary within broad limits (see Table
2). After rejection of several extreme values, the average
$[Fe/H]$ = -\,1.59.

\begin{table} [!h]
\begin{center}
{Table 2. The determinations of metallicity of NGC\,7006\\}
\vspace{0.2cm}
\begin{tabular}{|l|c|l|}
\hline
 $[Fe/H]$& method & Reference\\ \hline
 -2.0 & Washington photometry & Canterna (1975)\\
 -1.6 & DDO photometry from spectral scans & Hesser et al.(1977)\\
 -1.45 &low-resolution spectra & Searle and Zinn (1978)\\
 -1.2 & synthetic spectra &Bell and Gustafsson (1978)\\
 -1.53  & Q39 & Zinn (1980)\\
 -1.59  & compilation & Webbink (1981)\\
 -1.9 & spectral scans  & McClure and Hesser (1981)\\
 -1.45  & low-resolution spectra & Friel et al.(1982)\\
 -1.5 &low-resolution spectra & Cohen and Frogel (1982)\\
 -1.59  & compilation  & Zinn and West (1984)\\
 -1.77& $(B-V)_{o,g}$ & Buonanno et al.(1991)\\
 -1.63  &compilation & Harris (1996)\\
 -1.6  &medium-resolution spectra & Wachter et al.(1998)\\
 -1.55 &high-resolution spectra & Kraft et al. (1998)\\
 -1.78 &$V vs.(B-R)$ diagram  &  this work\\
  \hline

\end{tabular}
\end{center}
\end{table}

Intense studies of NGC\,7006 carried out by Sandage and Wildey (1967)
showed that its HB morphology does not correspond to its metallicity.
Later on, an entire group of intermediate-metallicity clusters was
found, in which HB appears to be redder than it should be for a given
metallicity; this discrepancy, known as "the problem of the second
parameter", has still not explained.

At the beginning of the 90-ies, in the study of Buonanno et al.(1991)
it was assumed that the HB color tends to be bluer with a decrease in
the distance from the cluster center. In our study (Gerashchenko, 2007),
it was proved for distances from the center of a cluster up to
$r\geq15^{\prime\prime}$.

In addition to that, in the above study, using the relations
obtained in the study of Ferraro et al.(1999) between metallicity
and GB parameters derived from the $V vs. (B-V)$ diagram, we
determined the metallicity and interstellar absorption for the
cluster. The metallicity (-\,1.62 in the Carretta and Gretton
(1997) system) corresponds to the averaged over earlier
determinations; however, the color excess ($E_{B-V}=0.^{m}15$)
exceeds most of previously found values ($E_{B-V}=0.^{m}05$).

In our study, we determine the HB parameters from the $V vs.(B-R)$
diagram, from which, using the formula (11 - 17), we calculate
metallicity and reddening of the cluster according to two
procedures, both including iterative approximations.

One of these procedures was presented in the study by Ferraro et
al. (1999), where the initial metallicity is estimated from the
formulae (11) and (12) and the interstellar absorption from the
ratio $E_{B-R}$ = $(B-R)_g$-$(B-R)_{o,g}$ and the formulae (13)
for $(B-R)_{o,g}$. These values are then improved by iterative
calculations according to the formulae (14-17) until the point
when two consecutive estimations for $[Fe/H]$ and $E_{B-R}$ differ
less than by 0.1dex, or $0.^{m}02$, respectively.

The other procedure was suggested by Sarajedini(1994). The system
of equations (10-13) is solved iteratively for $E_{B-R}$ =
$E_{0,(B-R)}$+$i\Delta E_{B-R}$, until the point when the
difference in metallicity determined from $\Delta V$ and
$(B-R)_{o, g}$ becomes smaller than 0.1dex. Here, $E_0,{B-R)}$ is
the initial value of the color excess $E_{B-R}$, $\Delta\,
E_{B-R}=0.1$ is the step of iterations, i - the number of them.

Within the errors of determination, both procedures yield
identical results: $E_{B-R} = 0.^{m}25 \pm0.^{m}02, [Fe/H] =
-\,1.78 \pm0.11$; the convergence of the results is reached very
rapidly.

\section {Conclusion}

Here, we have derived relations between GB parameters in the $V
vs. (B-R)$ diagrams for a globular cluster and its metallicity,
which make it possible to determine simultaneously the metallicity
and interstellar absorption for the cluster by iterative
calculations. The accuracy of these determinations appears to be
not worse than that reached by a similar method using $V vs.(B-V)$
diagrams: $\pm0.1$dex for the metallicity and $ \pm0.^{m}02 $ for
the color excess. The application of this technique to $BVR$
photometry of the cluster NGC\,7006 yields $[Fe/H]$ = - \, 1.78
and $E_{B-R} = 0.^{m}25$. The obtained metallicity is consistent
with that previously published within the limits of determination
errors. The color excess $E_{B-R}$, after the transformation into
$E_{B-V}$, confirms our previous determination $E_{B-V} =
0.^{m}15$ (Gerashchenko, 2007)

\section {Acknowledgements}
The author is grateful to N.N.Samus' for providing photometric
data for three clusters, to V.N.Frolov for useful discussion, and to
J.K.Ananjevskaya for her assistance in the preparation of this paper.

\end{document}